\documentclass[]{revtex4}
\usepackage{hyperref}
 
\begin{document}

\title{Gaugeon formalism in the framework of generalized BRST symmetry  }

\author{ Sudhaker Upadhyay}
 \email {  sudhakerupadhyay@gmail.com; 
 sudhaker@boson.bose.res.in}

\affiliation { S. N. Bose National Centre for Basic Sciences,\\
Block JD, Sector III, Salt Lake, Kolkata -700098, India. }
 \author{ Bhabani Prasad Mandal}
 \email {  bhabani.mandal@gmail.com} 
 
\affiliation { Department of Physics, Banaras Hindu University,  \\
Varanasi-221005, India.}

\begin{abstract}
We consider gaugeon formulations which discuss the quantum gauge freedom   covariantly  in the framework of generalized BRST transformation for the
Yang-Mills theory as well as the BRST invariant Higgs model.
  We generalize the
BRST symmetries of both the Yang-Mills theory and the Higgs model 
by making the transformation parameter finite and field-dependent.
Remarkably, we observe that the gaugeon Lagrangian which describe the 
quantum gauge freedom appear automatically in the effective  
 theories along with the natural shift in gauge parameters under
specific finite field-dependent parameters.

  \end{abstract}
\maketitle

\section{  Introduction}

In usual canonical quantization of the gauge theories there exits no gauge freedom at the quantum level as quantum theory is defined
only after gauge fixing. The gaugeon formulation provides a wider framework in which one can consider quantum gauge transformation
among a family of linear covariant gauges \cite{yo0,yok,yoko,yo1,yo2,yo3}. In this formulation the
so-called gaugeon fields are introduced as the quantum gauge freedom and such 
a formulation was originally proposed to restore the problem of gauge parameter renormalization. The shift of gauge parameter
which arises  through renormalization  \cite{haya} is naturally derived in this formulation by connecting theories in two different gauges within the
same family by a $q$-number gauge transformation \cite{yo0}. The main drawback of this formulation is the unphysical gaugeon fields which do not
contribute in the physical process. Thus, it is necessary to remove the gaugeon  modes by imposing subsidiary conditions. Initially  it was performed
by putting Gupta-Bleuler type of restriction which has its own limitations. Gaugeon formulation 
is  improved in certain cases by further extending the configuration space   to incorporate
the BRST quartet for gaugeon fields \cite{ki,mk} 
where Gupta-Bleuler type subsidiary condition is replaced by Kugo-Ojima type restrictions \cite{kugo, kugo1}.  The geugeon formalism with and without BRST symmetry have been studied  in many different contexts in quantum field theory
\cite{ki, mk, mk1, naka, rko, miu, mir1, mir2,sud0} as well as in the perturbative gravity \cite{sud}.

In the present work we would like to consider gaugeon formulation in the light of generalized BRST transformation \cite{sdj}.  The generalized BRST symmetry  by making the infinitesimal 
parameter finite and field-dependent is known as finite field-dependent
BRST (FFBRST) transformation and
 has been found many implications
in gauge theories \cite{sdj,sb,sb2,sdj1,rb,susk,jog,sb1,smm,fs,sud1,rbs,sudha}.  
This provides us sufficient motivations  to analyse the gaugeon formulation 
through generalized BRST symmetry.  For this purpose, we consider two different models, (i) the Yang-Mills theory \cite{ym}, the backbone of all frontier
high energy physics  and (ii) the
Higgs model \cite{higg,kibble} which provides a general framework
to explain the observed masses of the gauge bosons
by means of charged and neutral Goldstone bosons that end
up as the longitudinal components of the gauge bosons,  to describe the quantum gauge symmetry 
in the framework of gaugeon formalism.  

We extend the effective
action by introducing two gaugeon fields in both the models. 
Such an extended theory posses the quantum gauge symmetry
under which Lagrangian  remains form invariant. These gaugeon fields
do not contribute in physical processes and therefore lead
to unphysical gaugeon modes.
To remove the unphysical modes, we put a Gupta-Bleuler type 
condition on gaugeon fields which finds certain limitations.
This situation is further improved in the Higgs model  by extending the action by introducing the Faddeev-Popov ghosts associated with gaugeon fields.
Such an action
remains invariant under both the extended BRST symmetry and the extended quantum gauge
symmetry.
Now, we generalized the full BRST symmetries of   the theories
 by allowing the infinitesimal parameter to be finite and field-dependent
 with continuous interpolation of arbitrary parameter $\kappa$.
Further, we calculate the Jacobian of path integral measure in each case
for a specific finite field-dependent parameter and  show that the
Jacobian produces the exact gaugeon part to the  effective action
with renormalized gauge parameters. Therefore,
we claim that FFBRST transformation  with an appropriate 
transformation parameter produces the gaugeon effective action
having accurate shift in gauge parameter 
to describe the quantum gauge freedom. 
 Even though  we establish these results 
 with the help of two different but typical models, namely,   Yang-Mills theory and  Higgs model, these results hold 
 good for any arbitrary models in gaugeon formulation.

The paper is organized as follows.
In section II, we  discuss the preliminaries about FFBRST transformation.
Section III is devoted to study the gaugeon formalism of both the Yang-Mills
as well as of Higgs model. Within this section, we also
investigate the FFBRST transformation  and emergence of
gaugeon mode through Jacobian of path integral measure.
The conclusions are drawn in the last section.

\section{ The generalized BRST transformation: preliminaries}
In this section, we recapitulate the
generalized BRST transformation with finite field-dependent
parameter which is also
 known as FFBRST transformation\cite{sdj}.
 For this purpose, let us begin with the usual BRST transformation defined by
 \begin{equation}
\delta_b \phi = s_b\phi\ \eta,
\end{equation}
  where $s_b\phi$ is Slavnov variation of generic field $\phi$.
 
 The properties of the usual BRST transformation    do not depend on whether 
the transformation parameter $\eta$ whether it is finite or infinitesimal
and field-dependent or not, however, it must be anticommuting and space-time independent. These inspections give the freedom to 
generalize the BRST transformation by making the parameter, $\eta$ finite and field-dependent without
 affecting its properties.   To generalize the BRST transformation we 
start 
by making the  infinitesimal parameter field-dependent with introduction of an arbitrary parameter $\kappa\ 
(0\leq \kappa\leq 1)$.
We allow the fields, $\phi(x,\kappa)$, to depend on  $\kappa$  in such a way that $\phi(x,\kappa =0)=\phi(x)$ and $\phi(x,\kappa 
=1)=\phi^\prime(x)$, the transformed field.
Furthermore,  the  usual infinitesimal BRST transformation is defined generically 
as \cite{sdj}
\begin{equation}
{d\phi(x,\kappa)}=s_{b} [\phi (x) ]\Theta^\prime [\phi (x,\kappa ) ]{d\kappa}
\label{diff}
\end{equation}
where the $\Theta^\prime [\phi (x,\kappa ) ]{d\kappa}$ is the infinitesimal but field-dependent parameter.
The FFBRST transformation with the finite field-dependent parameter then can be 
constructed by integrating such infinitesimal transformation from $\kappa =0$ to $\kappa= 1$, to obtain
\begin{equation}
\phi^\prime\equiv \phi (x,\kappa =1)=\phi(x,\kappa=0)+s_b[\phi(x) ]\Theta[\phi(x) ]
\label{kdep}
\end{equation}
where 
\begin{equation}
\Theta[\phi(x)]=\int_0^1 d\kappa^\prime\Theta^\prime [\phi(x,\kappa^\prime)],\label{infi}
\end{equation}
 is the finite field-dependent parameter \cite{sdj}. 
This FFBRST transformation  is the symmetry  of the effective action. However, 
being transformation parameter field-dependent the 
path integral measure is no more invariant under such transformation.
The Jacobian of path integral measure changes non-trivially
under FFBRST transformation. To estimate the
Jacobian $J(\kappa)$  of the path integral measure $({\cal D}\phi)$  under FFBRST
 transformations   for a 
particular choices of the finite field-dependent parameter, $\Theta[\phi(x)]$,  
we first calculate the infinitesimal change  in Jacobian
using Taylor expansion as follows \cite{sdj}
\begin{equation}
\frac{1}{J}\frac{dJ}{d\kappa}=-\int d^4y\left [\pm s_b \phi (y,\kappa )\frac{
\partial\Theta^\prime [\phi (y,\kappa )]}{\partial\phi (y,\kappa )}\right],\label{jac}
\end{equation}
where $\pm$ sign refers to whether $\phi$ is a bosonic or a fermionic field.

Further, the Jacobian  $J(\kappa )$ can be replaced (within the functional integral) by
\begin{equation}
J(\kappa )\rightarrow \exp[iS_1[\phi(x,\kappa) ]],
\end{equation}
without changing the theory defined by the action $S_{eff}$ numerically   if and only if the following essential condition is satisfied \cite{sdj}
\begin{eqnarray}
 \int {\cal{D}}\phi (x) \;  \left [ \frac{1}{J}\frac{dJ}{d\kappa}-i\frac
{dS_1[\phi (x,\kappa )]}{d\kappa}\right ]\exp{[i(S_{eff}+S_1)]}=0. \label{mcond}
\end{eqnarray}
where $ S_1[\phi ]$ refers to some local functional of fields.

Consequently,  the functional  $S_1$ within functional integral
accumulate to
effective action $S_{eff}$ and therefore effective action modifies to $S_{eff}+S_1$ which becomes an 
extended effective 
action.
Hence, FFBRST transformation with an appropriate parameter $\Theta$ 
extends the effective action of the theory.
We utilise this fact to show that the gaugeon modes in the effective theory
which describes the quantum gauge freedom are generated through FFBRST transformation.
To produce the extra piece  $S_1$  in the effective action having   some extra fields  through
Jacobian calculation, we first insert a well-defined path integral measure corresponding to that 
extra fields  in the functional integral by hand before performing FFBRST transformation and
thereafter  Jacobian factor compensates the divergence factor. 
However, if the  extra piece  $S_1$ contains only original fields then we don't
need   any extra path integral measure before the performing FFBRST transformation.
\section{Gaugeon formalism and its emergence through generalized BRST symmetry}
In this section, we review the
Yokoyama gaugeon formalism to discuss the quantum gauge freedom for the Yang-Mills theory as well as for the Higgs model.
\subsection{BRST symmetric Yokoyama-Yang-Mills theory}
To analyse the gaugeon formalism for Yang-Mills theory, let us start with the effective Lagrangian density for four dimensional Yang-Mills theory in Landau gauge  
\begin{eqnarray}
{\cal L}_{YM} =-\frac{1}{4}F^a_{\mu\nu}F^{ \mu\nu a} -A^{a }_\mu\partial^\mu B^a
+i\partial^\mu c^a_\star D^{ab}_\mu c^b, \label{ym}
\end{eqnarray}
where $A_\mu^a$, $B^a$, $c^a$ and $c^a_\star$
are gauge field, multiplier field, ghost field and anti-ghost field respectively. Here field-strength tensor ($F^a_{\mu\nu}$) and covariant derivative ($D^{ab}_\mu$)
 are defined  by 
\begin{eqnarray}
F^a_{\mu\nu} &=&\partial_\mu A^a_\nu -\partial_\nu A^a_\mu +gf^{abc}A^b_\mu  A^c_\nu,\nonumber\\
D^{ab}_\mu &=& \partial_\mu\delta^{ab} -gf^{abc} A^c_\mu,
\end{eqnarray}
with   coupling constant $g$.
The Lagrangian density (\ref{ym}) is invariant under following nilpotent BRST transformations:
 \begin{eqnarray}
 \delta_b A^a_\mu &=& -D^{ab}_\mu c^b\ \eta,\  \
 \delta_b c^a=-\frac{g}{2} f^{abc}c^b c^c\ \eta,\nonumber\\
 \delta_b c^a_\star &=&-iB^a \  \eta,\ \ \ \ \
 \delta_b B^a=0,\label{BRS}
 \end{eqnarray}
where $\eta$ is an infinitesimal, anticommuting and global parameter.
Now, by introducing the gaugeon field $Y$ and its associated field $Y_\star$ subject
to the Bose-Einstein statistics, the Yokoyama  Lagrangian density for the Yang-Mills theory is demonstrated   as \cite{yok}
\begin{eqnarray}
{\cal L}_{Y}(\phi, \alpha) =-\frac{1}{4}F^a_{\mu\nu}F^{ \mu\nu a} -A^{a}_\mu\nabla^\mu B^a
-\partial_\mu Y_\star \partial^\mu Y +\frac{\varepsilon}{2} (Y_\star +\alpha^a B^a)^2
+i\nabla^\mu c^a_\star D^{ab}_\mu c^b,\label{ga}
\end{eqnarray}
where $\alpha^a$ is the group vector valued gauge-fixing parameter and $\varepsilon (\pm)$
is the sign factor. Here $\nabla_\mu$ refers the \textit{form covariant derivative}
 defined as
\begin{eqnarray}
\nabla_\mu V^a=\partial_\mu  V^a+gf^{abc} \alpha^b V^c\partial_\mu Y,\ \ \ \ (V^a=B^a, c^a_\star).
\end{eqnarray}
The Lagrangian density (\ref{ga}) remains invariant under following  BRST transformation: 
\begin{eqnarray}
 \delta_b A^a_\mu &=& -D^{ab}_\mu c^b\ \eta,\  \
 \delta_b c^a=-\frac{g}{2} f^{abc}c^b c^c\ \eta,\nonumber\\
 \delta_b c^a_\star &=&-iB^a \  \eta,\ \ 
 \delta_b B^a=0,\ \ 
 \delta_b Y=0,  \ \delta_b Y_\star =0.\label{brst}
\end{eqnarray}
Now, we define the following quantum gauge transformation under which  the Lagrangian density  (\ref{ga}) remains form invariant  \cite{yok}:
\begin{eqnarray}
&&A^a_\mu\longrightarrow \hat A^a_\mu = A^a_\mu +\tau (\alpha^a\partial_\mu Y +g 
f^{abc}A^b_\mu 
 \alpha^c Y),\nonumber\\
 &&B^a\longrightarrow \hat B^a = B^a +\tau  g f^{abc}B^b \alpha^c Y,\nonumber\\
 &&Y_\star\longrightarrow \hat Y_\star = Y_\star -\tau  \alpha^a B^a,\nonumber\\
 &&Y\longrightarrow \hat Y =Y,\nonumber\\
  &&c^a\longrightarrow \hat c^a =c^a +\tau g f^{abc} c^b \alpha^c Y,\nonumber\\
  &&c^a_\star\longrightarrow \hat c^a_\star =c^a_\star +\tau g f^{abc}c^b_\star 
 \alpha^c  Y, \label{quan}
\end{eqnarray}
where $\tau$ is an infinitesimal transformation parameter having bosonic nature.
The form invariance of Lagrangian density (\ref{ga}) under quantum gauge transformation (\ref{quan})
leads to the following  shift   in parameter:
\begin{equation}
\alpha^a\longrightarrow \hat\alpha^a =\alpha^a+\tau \alpha^a.\label{alp}
\end{equation}
Further, to remove the unphysical modes of the theory and to define physical states we  impose  
two  subsidiary conditions \cite{yok}
 \begin{eqnarray}
 Q_b|\mbox{phys}\rangle &=&0,\nonumber\\
 (Y_\star +\alpha^a B^a)^{(+)}|\mbox{phys}\rangle &=&0,\label{con}
 \end{eqnarray}
where $Q_b$ is the BRST charge calculated as
\begin{eqnarray}
Q =\int d^3x \left[-F^{ 0\nu a} D^{ab}_\nu c^b -i\frac{g}{2}f^{abc} \dot c^a_\star  c^b c^c 
- D^{0ab} c^b B^a \right].
\end{eqnarray}
 The   Kugo-Ojima type subsidiary condition  [first of Eq. (\ref{con})] is 
subjected 
to remove  the unphysical gauge field
modes from the total Fock space. However, the second Gupta-Bleuler
type condition guarantees that no gaugeon appears in the physical states. 
The second subsidiary condition is well-defined when
the combination $(Y_\star +\alpha^a B^a)$   satisfies
the following free equation \cite{yok}
\begin{eqnarray}
\partial_\mu \partial^\mu (Y_\star +\alpha^a B^a)  =0.
\end{eqnarray}
The above free equation assures the decomposition of  $(Y_\star +\alpha^a B^a)$ in positive and negative frequency parts. The subsidiary conditions (\ref{con} guarantee
the metric of our physical state-vector space to be positive semi-definite
\begin{equation}
\langle \mbox{phys}| \mbox{phys}\rangle\geq 0,
\end{equation} 
and consequently, we have a desirable physical subspace  on which our unitary physical
$S$-matrix exists.

 Now, we analyse the emergence of
 gaueon mode in the effective   Yang-Mills theory by calculating the Jacobian of 
 path integral measure under FFBRST transformation. 
First of all, we  construct the FFBRST transformation by making the 
infinitesimal parameter $\eta$ of Eq. (\ref{brst}) finite and
field-dependent (in the same fashion  as discussed in earlier section) as follows:
 \begin{eqnarray}
 \delta_b A^a_\mu &=& -D^{ab}_\mu c^b\ \Theta [\phi],\  \
 \delta_b c^a=-\frac{g}{2} f^{abc}c^bc^c\ \Theta[\phi],\nonumber\\
 \delta_b c^a_\star &=&-iB^a \  \Theta[\phi],\ \  
 \delta_b B^a=0,\ \
  \delta_b Y^a =0,\ \ 
 \delta_b Y^a_\star =0.
 \end{eqnarray}
 where $\Theta[\phi]$ is an arbitrary finite field-dependent parameter with ghost number $-1$.
Now, we choose the
following infinitesimal field-dependent parameter
 \begin{eqnarray}
 \Theta' [\phi] &=&\int d^4 y \left[ gf^{abc}\hat{\alpha}^b c_\star^c A_\mu^a\partial^\mu Y-
 \varepsilon \hat{\alpha}^ac_\star^a\left(\frac{1}{2}\hat{\alpha}^b B^b+Y_\star\right)+c_\star^a 
 B^a (B^b)^{-2}\partial_\mu Y_\star\partial^\mu Y\right.\nonumber\\
 &-&\left.\frac{\varepsilon}{2} c_\star^a  B^a (B^b)^{-2} Y^2_\star\right],
 \end{eqnarray}
to construct the specific $\Theta[\phi]$  using relation (\ref{infi}), where $\hat\alpha$ denotes the shifted gauge parameter as defined  in  (\ref{alp}).
 Now, exploiting the relation
 (\ref{jac}),  the infinitesimal change in Jacobian   for above $\Theta'[\phi]$ yields
 \begin{eqnarray}
 \frac{1}{J(\kappa)}\frac{dJ(\kappa)}{d\kappa}&=& -\int d^4 x\left[gf^{abc}\hat{\alpha}^b (iB^c) A_\mu^a\partial^\mu Y-gf^{abc}\hat{\alpha}^b(D_\mu c)^a c_\star^c\partial^\mu Y-
 \varepsilon \hat{\alpha}^a(iB^a)\left(\frac{1}{2}\hat{\alpha}^b B^b+Y_\star\right)\right.\nonumber\\
 &+&\left. i\partial_\mu Y_\star\partial^\mu Y -i\frac{\varepsilon}{2}Y^2_\star\right],\nonumber\\
 &=&  -\int d^4 x\left[igf^{abc}\hat{\alpha}^b  B^c  A_\mu^a\partial^\mu 
 Y+gf^{abc}\hat{\alpha}^b c_\star^c\partial^\mu Y(D_\mu c)^a-
i\frac{\varepsilon}{2}  (\hat{\alpha}^a B^a)^2+ i\varepsilon\hat{\alpha}^a B^a Y_\star\right.\nonumber\\
 &+&\left. i\partial_\mu Y_\star\partial^\mu Y -i\frac{\varepsilon}{2}Y^2_\star\right].\label{ja}
 \end{eqnarray}
 Now,  the Jacobian $J$ can be written properly in terms of local fields as $e^{iS_1}$ 
for the following assumption
 \begin{eqnarray}
 S_1 [\phi(x,\kappa), \kappa] &=&\int d^4 x \left[\xi_1 (\kappa) gf^{abc}\hat{\alpha}^b  B^c  
 A_\mu^a\partial^\mu 
 Y +  \xi_2 (\kappa) gf^{abc}\hat{\alpha}^b c_\star^c\partial^\mu Y(D_\mu c)^a+\xi_3 (\kappa)
  (\hat{\alpha}^a B^a)^2\right.\nonumber\\
  &+&\left. \xi_4 (\kappa)\hat{\alpha}^a B^aY_\star +\xi_5 (\kappa) \partial_\mu Y_\star\partial^\mu Y +\xi_6 (\kappa) Y^2_\star \right],\label{s22}
 \end{eqnarray} 
 where $\xi_i (i=1,2,..,6)$ are arbitrary $\kappa$-dependent constants
 satisfying following initial
 boundary conditions
 \begin{equation}
 \xi_i (\kappa=0)=0.\label{ini}
 \end{equation}
At the physical ground, the theory remains unaltered when the above $S_1$ and 
 change in Jacobian given in (\ref{ja}) satisfy the crucial condition (\ref{mcond}).
To check this consistency,  we first calculate 
 the infinitesimal difference in $S_1$ with respect to parameter $\kappa$ with the help of (\ref{diff}) as follows
  \begin{eqnarray}
 \frac{S_1 [\phi(x,\kappa), \kappa]}{d\kappa}&=&\int d^4 x \left[
 \frac{d\xi_1}{d\kappa}   gf^{abc}\hat{\alpha}^b  B^c  
 A_\mu^a\partial^\mu 
 Y + \frac{d\xi_2}{d\kappa}  gf^{abc}\hat{\alpha}^b c_\star^c\partial^\mu Y(D_\mu c)^a+
 \frac{d\xi_3}{d\kappa} 
  (\hat{\alpha}^a B^a)^2\right.\nonumber\\
  &+&\left. \frac{d\xi_4}{d\kappa} \hat{\alpha}^a B^aY_\star +\frac{d\xi_5}{d\kappa}  \partial_\mu Y_\star\partial^\mu Y +\frac{d\xi_6}{d\kappa}  Y^2_\star -\xi_1 (\kappa) gf^{abc}\hat{\alpha}^b  B^c  
(D_\mu c)^a\partial^\mu  Y \Theta'\right.\nonumber\\
&-&\left.  \xi_2 (\kappa) gf^{abc}\hat{\alpha}^b (iB)^c\Theta' \partial^\mu Y(D_\mu c)^a
\right].\label{s1}
 \end{eqnarray}
Now, the consistency condition   (\ref{mcond}) together with
Eqs. (\ref{ja}) and (\ref{s1}) leads
 \begin{eqnarray}
&& \int d^4 x \left[
 \left(\frac{d\xi_1}{d\kappa} +1\right) gf^{abc}\hat{\alpha}^b  B^c  
 A_\mu^a\partial^\mu 
 Y +  \left(\frac{d\xi_2}{d\kappa}-i\right)  gf^{abc}\hat{\alpha}^b c_\star^c\partial^\mu Y(D_\mu c)^a\right.\nonumber\\
 &+&\left.\left(
 \frac{d\xi_3}{d\kappa}-1\right) 
  (\hat{\alpha}^a B^a)^2+ \left(\frac{d\xi_4}{d\kappa}-\varepsilon\right) \hat{\alpha}^a B^aY_\star +\left(\frac{d\xi_5}{d\kappa} +1\right) \partial_\mu Y_\star\partial^\mu Y  \right.\nonumber\\
&+&\left. \left( \frac{d\xi_6}{d\kappa}  -\frac{\varepsilon}{2}\right) Y^2_\star -(\xi_1  -i\xi_2 ) gf^{abc}\hat{\alpha}^b  B^c  
(D_\mu c)^a\partial^\mu  Y \Theta' 
\right] =0,\label{mco}
 \end{eqnarray}
 where the non-local ($\Theta'$ dependent) term vanishes leading to following constraint:
 \begin{equation}
   \xi_1 (\kappa)-i\xi_2 (\kappa) =0.
 \end{equation}
However, the disappearance of local terms  from the LHS of expression (\ref{mco})
 leads to following exactly solvable linear differential equations 
 \begin{eqnarray}
\frac{d\xi_1}{d\kappa} +1 &=&0,\ \  
 \frac{d\xi_2}{d\kappa}-i=0,\nonumber\\
\frac{d\xi_3}{d\kappa}-1 &=&0,\ \
\frac{d\xi_4}{d\kappa} -\varepsilon=0,\nonumber\\
\frac{d\xi_5}{d\kappa}+1 &=&0,\ \ \
\frac{d\xi_6}{d\kappa} -\frac{\varepsilon}{2}=0.
 \end{eqnarray}  
  The solutions of above equations satisfying the initial boundary conditions (\ref{ini})  are 
 \begin{eqnarray}
 \xi_1 (\kappa) =-\kappa,\ \ \ \xi_2(\kappa) =i\kappa,\
  \ \xi_3(\kappa) = +\kappa,
 \ \ \xi_4(\kappa) =+\varepsilon\kappa,\
  \ \xi_5(\kappa) = -\kappa, \ \ \xi_6(\kappa) =\frac{\varepsilon}{2}\kappa.  
 \end{eqnarray}
With these solutions, the expression (\ref{s22}) at $\kappa =1$ receives the following form:

 \begin{eqnarray}
 S_1 [\phi(x,1), 1]&=&\int d^4 x \left[- gf^{abc}\hat{\alpha}^b  B^c  
 A_\mu^a\partial^\mu 
 Y + i gf^{abc}\hat{\alpha}^b c_\star^c\partial^\mu Y(D_\mu c)^a+ 
  (\hat{\alpha}^a B^a)^2\right.\nonumber\\
  &+&\left. \varepsilon \hat{\alpha}^a B^aY_\star - \partial_\mu Y_\star\partial^\mu Y +\frac{\varepsilon}{2} Y^2_\star \right].
 \end{eqnarray}
 Now, by adding $S_1 [\phi(x,1), 1]$ to the effective action corresponding to (\ref{ym}), we get
\begin{eqnarray}
\int d^4x\ {\cal L}_{YM} + S_1 [\phi(x,1), 1] &=&
\int d^4x\ \left[-\frac{1}{4}F^a_{\mu\nu}F^{ \mu\nu a} -A^{a}_\mu\nabla^\mu B^a
-\partial_\mu Y_\star \partial^\mu Y  \right.\nonumber\\
&+&\left.\frac{\varepsilon}{2} (Y_\star +\hat\alpha^a B^a)^2
+i\nabla^\mu c^a_\star D^{ab}_\mu c^b  \right],\nonumber\\
&=& \int d^4x\ {\cal L}_{Y}(\phi, \hat{\alpha}),
\end{eqnarray}
which is nothing but the  gaugeon action for Yang-Mills theory
 with shifted gauge parameter $\hat \alpha^a=\alpha^a(1+\tau)$.
Hence, we end up the subsection with following remark that 
under specific generalized BRST transformation the   gaugeon modes in the effective Yang-Mills action appears manifestly.

\subsection{BRST symmetric Higgs model}
To describe the gaugeon formulation of Higgs model in the framework of FFBRST transformation, we
begin with the classical Lagrangian density of the Higgs model defined by
 \begin{eqnarray}
 {\cal L}_H =-\frac{1}{4}F_{\mu\nu}F^{\mu\nu} +(D_\mu \varphi)^\dag (D^\mu\varphi) +\mu^2
 \varphi^\dag \varphi -\frac{\lambda}{2} (\varphi^\dag \varphi)^2,\label{la}
 \end{eqnarray}
 where $\varphi$ is the complex scalar field, $\mu^2$ and $\lambda$ are positive constants . 
 The
 field-strength tensor and  covariant derivative are defined, respectively, by 
 \begin{eqnarray}
 F_{\mu\nu} &=&\partial_\mu A_\nu
 -\partial_\nu A_\mu,\nonumber\\
 D_\mu &=& \partial_\mu -eA_\mu.
\end{eqnarray}  
Here the complex scalar field $\varphi$  has the following  vacuum expectation value: 
\begin{equation}
\langle0|\varphi|0\rangle =\frac{v}{\sqrt{2}} = \frac{\mu}{\sqrt{\lambda}}.
\end{equation}
It is well known that the Lagrangian density   (\ref{la}) is gauge invariant. 
Therefore, to quantize it correctly we need to break the local gauge invariance by
fixing a suitable gauge. 
There are many choices for gauge-fixing condition. For example,
the  gauge-fixed Lagrangian density corresponding to $R_\xi$ gauge condition
introduced by Fujikawa, Lee and
Sanda \cite{fuz} is given by
\begin{eqnarray}
{\cal L}_{gf} = \frac{1}{2\xi}B^2 +B  \left( \partial_\mu A^\mu +\frac{1}{\xi} M\chi\right),\label{gf}
\end{eqnarray}
  where $B$ is multiplier field and $\xi$ is a numerical gauge-fixing parameter. here $M=ev$ is the mass of $A_\mu$ acquired through the spontaneous symmetry breaking and hermitian field $\chi$ is the Goldstone mode defined  along with physical Higgs mode $\psi$  as 
  \begin{equation}
\varphi =\frac{1}{ \sqrt{2}} (v+\psi +i\chi ).\label{par}
  \end{equation} 
 Now,  the Faddeev-Popov ghost term 
 corresponding to the
 above gauge-fixing term is constructed as 
 \begin{equation}
 {\cal L}_{gh} =-i\partial_\mu c_\star \partial^\mu c +ic_\star \frac{M^2}{\xi} c,\label{gh}
 \end{equation}
 where $c$ ad $c_\star$ are the ghost and anti-ghost fields respectively.
 
To analyse the gaugeon
formalism for the $R_\xi$ gauge avoiding  non-polynomial terms into the
Lagrangian density, we use the following parametrization  \cite{miu}
\begin{equation}
\varphi (x) =(v+\rho (x))e^{i\pi(x)/v},\label{par1}
\end{equation}
instead of   (\ref{par}).
Here fields $\rho$ and $\pi$ get resemblance with fields $\psi$ and $\chi$ of Eq. (\ref{par}).
In terms of the parametrization, the Lagrangian density given in Eq. (\ref{la})
is expressed as
\begin{eqnarray}
{\cal L}_H& = &-\frac{1}{4} F_{\mu\nu} F^{\mu\nu} +\frac{1}{2}M^2 \left(1+\frac{e}{M}\rho\right)^2 \left( A_\mu -\frac{1}{M}\partial_\mu\pi\right)^2\nonumber\\
&+&\frac{1}{2}(\partial_\mu\rho\partial^\mu\rho -\lambda v^2\rho^2) -\frac{1}{2}\lambda v
\rho^3 -\frac{\lambda}{8}\rho^4 +\frac{1}{8}\lambda
v^4,
\end{eqnarray}
where $\sqrt{\lambda}v$ is the mass of Higgs boson $\rho$.
The above Lagrangian density is invariant under
following classical gauge transformations:
\begin{eqnarray}
A_\mu (x)&\longrightarrow &A'_\mu (x) =A_\mu (x) +\partial_\mu \Lambda (x),\nonumber\\
\pi (x)&\longrightarrow &\pi' (x) =\pi (x) +M \Lambda (x),\nonumber\\
\rho (x)&\longrightarrow &\rho ' (x) =\rho (x),
\end{eqnarray}
where $\Lambda (x)$ is an arbitrary local parameter of transformation.
Now, we recast the gauge-fixing and ghost terms given in   (\ref{gf}) and (\ref{gh}) in accordance 
with parametrization (\ref{par1})  as follows:
\begin{eqnarray}
{\cal L}_{gf}& = &\frac{1}{2}\alpha_1 B^2 +B  \left( \partial_\mu A^\mu + \beta_1 M\pi\right),\nonumber\\
 {\cal L}_{gh} &=&-i\partial_\mu c_\star \partial^\mu c +i\beta_1 c_\star  {M^2}  c,
\end{eqnarray}
where $\alpha_1$ and $\beta_1$ are gauge parameters.

Now, the effective Lagrangian density for Higgs model, 
\begin{equation}
{\cal L}_{eff} ={\cal L}_{H} +{\cal L}_{gf}+{\cal L}_{gh},\label{effe}
\end{equation} 
possesses the following nilpotent BRST transformation:
\begin{eqnarray}
 \delta_b A_\mu &=& -\partial_\mu c \ \eta,\ \ \delta_b \pi =-Mc\ \eta,\nonumber\\
 \delta_b \rho &=& =0,\ \ \delta_b c =0,\nonumber\\
 \delta_b c_\star &=& iB  \  \eta,\ \ 
 \delta_b B=0,\label{br}
\end{eqnarray}
where $\eta$ is an anticommuting global parameter.

Further, to analyse the quantum gauge freedom of Higgs model we extend the effective 
Lagrangian density (\ref{effe}) to a most general gaugeon Lagrangian density by introducing the
gaugeon field $Y$ and its associated field $Y_\star$ 
as well as the corresponding ghost fields  $K $ and $K_\star$   as
\begin{eqnarray}
{\cal L}_{YH} &=& {\cal L}_{H}+ B\partial_\mu A^\mu -\partial_\mu Y_\star \partial^\mu Y  +(\beta_1 B +
\beta_3 Y_\star )M\pi \nonumber\\
&+& (\beta_2 B +\beta_4 Y_\star)M^2 Y +\frac{1}{2}\alpha_1 B^2 +\alpha_2 B Y_\star
+\frac{1}{2}\alpha_3 Y_\star ^3\nonumber\\
&-&i\partial_\mu c_\star \partial^\mu c -i\partial_\mu K_\star \partial^\mu 
K +i(\beta_1 c_\star +\beta_3 K_\star )M^2 c \nonumber\\
& +& i(\beta_2 c_\star +\beta_4 K_\star )M^2 K,\label{yh}
\end{eqnarray}
where $\alpha_i (i=2,3)$ and $\beta_i (i=2,3,4)$ are constant 
gauge parameters. 
Now, the gaugeon fields and respective ghost fields vary under the BRST transformation 
as follows:
\begin{eqnarray}
\delta_b Y &=&-K \  \eta,\ \ \
\delta_b K=0,\nonumber\\
\delta_b K_\star &=& iY_\star \  \eta,\ \
\delta_b Y_\star =0,\label{brs}
\end{eqnarray}
 and form the BRST quartet. 
The gaugeon Lagrangian density (\ref{yh}) is invariant under
 the effect of combined  BRST transformations  (\ref{br}) and (\ref{brs}).
Consequently, the corresponding BRST charge $Q_b$  annihilates the physical subspace of
${\cal V}_{phys}$ of total Hilbert space, i.e.
\begin{eqnarray}
Q_b|\mbox{phys}\rangle =0.
\end{eqnarray}
 This single subsidiary condition of Kugo-Ojima type removes both the
 unphysical gauge modes as well as unphysical gaugeon modes. 
 
The gaugeon Lagrangian density  (\ref{yh}) also admits the following quantum gauge transformations:
\begin{eqnarray}
 A_\mu &\longrightarrow & \hat A_\mu =A_\mu +\tau \partial_\mu Y,\nonumber\\ 
 \pi_\mu &\longrightarrow & \hat \pi =\pi +\tau M Y,\nonumber\\
 Y_\star &\longrightarrow & \hat Y_\star =Y_\star -\tau B,\nonumber\\
 B &\longrightarrow & \hat B =B,\nonumber\\
 Y &\longrightarrow & \hat Y= Y,\nonumber\\
  c &\longrightarrow & \hat c=c +\tau K,\nonumber\\
K_\star &\longrightarrow & \hat K_\star =K_\star -\tau c_\star,\nonumber\\ 
c_\star &\longrightarrow & \hat c_\star =c_\star,\nonumber\\ 
K &\longrightarrow & \hat K =K.\label{q}
 \end{eqnarray}
 Under the above quantum gauge transformation  ${\cal L}_{YH}$ remains form invariant
 leading to the following shift in gauge parameters:
 \begin{eqnarray}
 \alpha_1 \longrightarrow \hat{\alpha}_1 &=&\alpha_1 +2\alpha_2\tau +\alpha_3 \tau^2,\nonumber\\
  \alpha_2 \longrightarrow \hat {\alpha}_2 &=&\alpha_2 + \alpha_3\tau,\nonumber\\ 
 \alpha_3 \longrightarrow  \hat{\alpha}_3 &=&\alpha_3,\nonumber\\ 
\beta_1 \longrightarrow  \hat{\beta}_1   &=&\beta_1 +\beta_3\tau,\nonumber\\ 
\beta_2 \longrightarrow \hat {\beta}_2 &=&\beta_2 +(\beta_4 -\beta_1)\tau -\beta_3\tau^2,\nonumber\\
\beta_3 \longrightarrow \hat{\beta}_3 &=&\beta_3,\nonumber\\
\beta_4 \longrightarrow \hat{\beta}_4 &=&\beta_4 -\beta_3\tau.\label{p}
 \end{eqnarray} 
 We observe that the quantum gauge transformations   (\ref{q})  commute  with BRST transformations mentioned  in  (\ref{brs}). Consequently, it is confirmed that the Hilbert space spanned from
 physical states annihilated by BRST charge is also invariant under the quantum gauge
transformations.
 
Now, we analyse the emergence of
 gaueon mode in effective action for  Higgs model  by calculating the Jacobian of 
 path integral measure under FFBRST transformation. 
To achieve this goal, we  construct the FFBRST transformation  by making the 
infinitesimal parameter $\eta$ of  (\ref{br}) and (\ref{brs}) finite and
field-dependent such that
  \begin{eqnarray}
 \delta_b A_\mu &=& -\partial_\mu c \ \Theta [\phi],\ \ \delta_b \pi =-Mc\ \Theta[\phi],\nonumber\\
 \delta_b \rho &=& =0,\ \ \delta_b c =0,\nonumber\\
 \delta_b c_\star &=& iB  \  \Theta[\phi],\ \ 
 \delta_b B=0,\nonumber\\
  \delta_b Y &=&-K \  \Theta[\phi],\ \ \
 \delta_b K=0,\nonumber\\
  \delta_b K_\star &=& iY_\star \  \Theta[\phi],\ \
 \delta_b Y_\star =0,\label{ffbrs}
 \end{eqnarray}
 where $\Theta[\phi]$ is finite field-dependent parameter is constructed from
 following infinitesimal field-dependent parameter 
\begin{eqnarray}
 \Theta' [\phi]& =&  \int d^4 y \left[\partial_\mu K_\star \partial^\mu Y -\hat\beta_2 c_\star M^2Y -\hat\beta_3 K_\star M\pi -\hat\beta_4 K_\star M^2Y\right. \nonumber\\
 &-&\left.\frac{1}{2}c_\star (\hat{\alpha}  B+
 \hat{\alpha}_2 Y_\star) -\frac{1}{2}K_\star (\hat{\alpha}_2 B+\hat{\alpha}_3 Y_\star)  \right].
 \end{eqnarray}
Here $\hat {\alpha}_2,  \hat{\alpha}_3,  \hat{\beta}_1,  \hat {\beta}_2,
  \hat{\beta}_3$ and   $ \hat{\beta}_4 $  are the shifted gauge parameters having same definition as in (\ref{p}). 
However, parameter  $\hat{\alpha}$ is defined in terms of $\tau$ explicitly as
\begin{eqnarray}
 \hat{\alpha} = 2\alpha_2\tau +\alpha_3 \tau^2.
 \end{eqnarray} 
We again calculate the infinitesimal change in Jacobian
 for path integral measure under FFBRST transformation in the 
 same way as calculated in the last subsection
 \begin{eqnarray}
 \frac{1}{J(\kappa)}\frac{dJ(\kappa)}{d\kappa}&=& \int d^4x\left[-i
 \partial_\mu Y_\star \partial^\mu Y +K\partial_\mu \partial^\mu  K_\star +i\hat\beta_2 BM^2 Y +\hat\beta_2 K c_\star
 M^2 +i\hat\beta_3 Y_\star M\pi\right.\nonumber\\
 &+&\left. \hat\beta_3 M^2 cK_\star +i\hat\beta_4 Y_\star M^2 Y +\hat\beta_4 KK_\star M^2 
 +\frac{1}{2}iB(\hat\alpha  B +\hat\alpha_2 Y_\star)\right.\nonumber\\
 &+&\left. \frac{1}{2}iY_\star(\hat\alpha_2 B +\hat\alpha_3 Y_\star)\right].
 \end{eqnarray} 
Dropping the total derivative terms the above expression reduces to 
 \begin{eqnarray}
 \frac{1}{J(\kappa)}\frac{dJ(\kappa)}{d\kappa}&=& \int d^4x\left[-i
 \partial_\mu Y_\star \partial^\mu Y +\partial_\mu K_\star\partial^\mu K  +i\hat\beta_2 BM^2 Y -\hat\beta_2  c_\star
 M^2K +i\hat\beta_3 Y_\star M\pi\right.\nonumber\\
 &-&\left. \hat\beta_3 K_\star M^2 c +i\hat\beta_4 Y_\star M^2 Y -\hat\beta_4 K_\star M^2 K +\frac{1}{2}iB(\hat\alpha  B +\hat\alpha_2 Y_\star)\right.\nonumber\\
 &+&\left. \frac{1}{2}iY_\star(\hat\alpha_2 B +\hat\alpha_3 Y_\star)\right].\label{j2}
 \end{eqnarray}
 Further, by considering all the terms appearing in the above 
  expression, we postulate the functional $S_1$ to have following form:
 \begin{eqnarray}
 S_1 [\phi(x,\kappa), \kappa]&=&   \int d^4x\left[\xi_1(\kappa) 
 \partial_\mu Y_\star \partial^\mu Y +\xi_2(\kappa)\partial_\mu K_\star\partial^\mu K  +\xi_3(\kappa) \hat\beta_2 BM^2 Y +\xi_4(\kappa)\hat\beta_2  c_\star
 M^2K \right.\nonumber\\
 &+&\left. \xi_5 (\kappa)\hat\beta_3 Y_\star M\pi+ \xi_6(\kappa) \hat\beta_3 K_\star M^2 c +\xi_7(\kappa) \hat\beta_4 Y_\star M^2 Y +\xi_8(\kappa)\hat\beta_4 K_\star M^2 K
 \right.\nonumber\\
 &+&\left. \xi_9 (\kappa)B(\hat\alpha  B +\hat\alpha_2 Y_\star)+ \xi_{10}
 (\kappa)Y_\star(\hat\alpha_2 B +\hat\alpha_3 Y_\star)\right],
 \end{eqnarray}
 where all $\kappa$ dependent constant parameters $(\xi_i,  i=1,2,...,10)$
 are prescribed to satisfy the following initial boundary conditions
 \begin{eqnarray}
 \xi_i(\kappa=0)=0.\label{bound}
 \end{eqnarray}
Now,  the infinitesimal change in $S_1$ is evaluated as 
 \begin{eqnarray}
 \frac{dS_1}{d\kappa}&=&   \int d^4x\left[\frac{d\xi_1 }{d\kappa}
 \partial_\mu Y_\star \partial^\mu Y +\frac{d\xi_2 }{d\kappa}\partial_\mu K_\star\partial^\mu K  +\frac{d\xi_3}{d\kappa} \hat\beta_2 BM^2 Y +\frac{d\xi_4 }{d\kappa}\hat\beta_2  c_\star
 M^2K \right.\nonumber\\
 &+&\left. \frac{d\xi_5 }{d\kappa}\hat\beta_3 Y_\star M\pi+\frac{d\xi_6 }{d\kappa} \hat\beta_3 K_\star M^2 c +\frac{d\xi_7 }{d\kappa} \hat\beta_4 Y_\star M^2 Y +\frac{d\xi_8 }{d\kappa}\hat\beta_4 K_\star M^2 K
 \right.\nonumber\\
 &+&\left. \frac{d\xi_9 }{d\kappa}B(\hat\alpha  B +\hat\alpha_2 Y_\star)+ \frac{d\xi_{10} }{d\kappa}
 (\kappa)Y_\star(\hat\alpha_2 B +\hat\alpha_3 Y_\star)-(\xi_1 +i\xi_2 )\partial_\mu Y_\star \partial^\mu K\Theta '
  \right.\nonumber\\
 &-&\left.(\xi_3 +i\xi_4 )\hat\beta_2 BM^2 K\Theta '
 -(\xi_5 + i\xi_6 )\hat\beta_3 YM^2 c\Theta '-(\xi_7 + i\xi_8 )Y_\star M^2
  K\Theta'\right],\label{s2}
 \end{eqnarray}
 where we have utilized the relation (\ref{diff}).
 The essential condition (\ref{mcond}) which validates the functional $S_1$
 together with Eqs. (\ref{j2}) and (\ref{s2}) yields
 \begin{eqnarray}
&& \int d^4x\left[i\left(\frac{d\xi_1 }{d\kappa} +1 \right)
 \partial_\mu Y_\star \partial^\mu Y +\left(i\frac{d\xi_2 }{d\kappa}-1\right)\partial_\mu K_\star\partial^\mu K  +i\left(\frac{d\xi_3}{d\kappa}-1 \right)\hat\beta_2 BM^2 Y  \right.\nonumber\\
 &+&\left.\left(i\frac{d\xi_4 }{d\kappa}+1\right)\hat\beta_2  c_\star
 M^2K+\left(\frac{d\xi_5 }{d\kappa}-1\right)\hat\beta_3 Y_\star M\pi+\left(i\frac{d\xi_6 }{d\kappa}+1\right) \hat\beta_3 K_\star M^2 c \right.\nonumber\\
 &+&\left.\left(\frac{d\xi_7 }{d\kappa} -1\right)\hat\beta_4 Y_\star M^2 Y +\left(i\frac{d\xi_8 }{d\kappa}+1\right)\hat\beta_4 K_\star M^2 K + \left( \frac{d\xi_9 }{d\kappa}-\frac{1}{2}\right)iB(\hat\alpha  B +\hat\alpha_2 Y_\star)
 \right.\nonumber\\
 &+&\left. \left(\frac{d\xi_{10} }{d\kappa} -\frac{1}{2}\right)i
 Y_\star(\hat\alpha_2 B +\hat\alpha_3 Y_\star)-(\xi_1 +i\xi_2 )\partial_\mu Y_\star \partial^\mu K\Theta ' -(\xi_3 +i\xi_4 )\hat\beta_2 BM^2 K\Theta '
  \right.\nonumber\\
 &-&\left.(\xi_5 + i\xi_6 )\hat\beta_3 YM^2 c\Theta'
 -(\xi_7 + i\xi_8 )Y_\star M^2 K\Theta '\right]=0.
 \end{eqnarray}
 Comparing the coefficients of the various terms present  in the
 above expression from LHS to RHS, we get following differential 
 equations
 \begin{eqnarray}
 && \frac{d\xi_1 }{d\kappa} +1 =0,\ \ i\frac{d\xi_2 }{d\kappa}-1 =0,\nonumber\\
  && \frac{d\xi_3}{d\kappa}-1 =0,\ \ i\frac{d\xi_4 }{d\kappa}+1 =0,\nonumber\\ && \frac{d\xi_5 }{d\kappa}-1=0,\ \ i\frac{d\xi_6 }{d\kappa}+1=0,\nonumber\\ 
  && \frac{d\xi_7 }{d\kappa} -1 =0,\ \ i\frac{d\xi_8 }{d\kappa}+1=0,\nonumber\\ && \frac{d\xi_9 }{d\kappa}-\frac{1}{2} =0,\ \  \frac{d\xi_{10} }{d\kappa} -\frac{1}{2}=0,
  \end{eqnarray}  
  together with
  \begin{eqnarray}
     && \xi_1 +i\xi_2 =0,\ \ \xi_3 +i\xi_4  =0,\nonumber\\
  &&\xi_5 + i\xi_6  =0,\ \ \xi_7 + i\xi_8 =0.
 \end{eqnarray}
The solutions of above equations satisfying the initial conditions (\ref{bound}) are
\begin{eqnarray}
&&\xi_1 =-\kappa,\ \ \xi_2 =-i \kappa,\ \ \xi_3 = \kappa,\nonumber\\
&& \xi_4
= i \kappa,\ \ \xi_5 = \kappa,\ \ \xi_6 =i\kappa,\nonumber\\
&&\xi_7 = 
 \kappa,\ \ \xi_8 =i\kappa,\ \ \xi_9 = \frac{1}{2}\kappa,\nonumber\\
&& \xi_{10} =\frac{1}{2}\kappa.
\end{eqnarray}
With these identifications of constant parameters $\xi_i$, the
exact form of $S_1$ is given by 
 \begin{eqnarray}
 S_1 [\phi(x,\kappa), \kappa ]&=&   \int d^4x\left[- 
 \kappa\partial_\mu Y_\star \partial^\mu Y -i\kappa\partial_\mu K_\star\partial^\mu K  + \kappa\hat \beta_2 BM^2 Y +i\kappa\hat\beta_2  c_\star
 M^2K \right.\nonumber\\
 &+&\left.  \kappa\hat\beta_3 Y_\star M\pi+ i\kappa \hat\beta_3 K_\star M^2 c +  \kappa\hat\beta_4 Y_\star M^2 Y +i\kappa\hat\beta_4 K_\star M^2 K
 \right.\nonumber\\
 &+&\left. \frac{\kappa}{2}B(\hat\alpha  B +\hat\alpha_2 Y_\star)+\frac{\kappa}{2} Y_\star(\hat\alpha_2 B +\hat\alpha_3 Y_\star)\right],
 \end{eqnarray}
 which vanishes at $\kappa =0$.
 However, functional $S_1$ at $\kappa=1$ (under FFBRST transformation)  accumulate to the effective action (\ref{effe}) within functional integral as
 \begin{eqnarray}
 \int d^4 x\ {\cal L}_{eff} + S_1 [\phi(x,1),1]&=& \int d^4 x\left[
 {\cal L}_{H}+ B\partial_\mu A^\mu -\partial_\mu Y_\star \partial^\mu Y  +(\hat\beta_1 B +
\hat\beta_3 Y_\star )M\pi\right. \nonumber\\
&+& \left.(\hat\beta_2 B +\hat\beta_4 Y_\star)M^2 Y +\frac{1}{2}\hat\alpha_1 B^2 +\hat\alpha_2 B Y_\star
+\frac{1}{2}\hat\alpha_3 Y_\star ^3\right.\nonumber\\
&-&\left.i\partial_\mu c_\star \partial^\mu c -i\partial_\mu K_\star \partial^\mu 
K +i(\hat\beta_1 c_\star +\hat\beta_3 K_\star )M^2 c \right.\nonumber\\
& +&\left. i(\hat\beta_2 c_\star +\hat\beta_4 K_\star )M^2 K \right], 
 \end{eqnarray}
 which is nothing but the BRST invariant effective action for
 gaugeon Higgs model with shifted gauge parameters. 
 Therefore, we conclude that under generalized BRST transformations with appropriate
 finite field-dependent parameter the
 gaugeon modes to describe quantum gauge freedom in Higgs model appear naturally in the effective action. 
\section{Conclusions}

We have first evoked  the gaugeon formalism for both the
Yang-Mills theory \cite{mk1} and Higgs model \cite{miu}. Following the Refs. \cite{mk1,miu}, we have 
extended the configuration space by introducing the
gaugeon field and its associated field in the effective actions of these models. Further,    the  quantum gauge transformation has been derived for such  extended actions. 
Under quantum gauge transformation the extended action remains form invariant
along with
a shift in gauge parameters. These natural shift in gauge parameters get
resemblance with those which appear  through proper renormalization \cite{yo0}.
Since these gaugeon fields are unphysical  and therefore, one needs to remove them. For this purpose, we have inserted a subsidiary
condition of Gupta-Bleuler type for Yang-Mills theory which removes
the unphysical gaugeon modes. But the Gupta-Bleuler type restriction has 
certain limitations. This situation is improved  in the  Higgs model   where we have enlarged 
the configuration space  by incorporating
the ghost fields corresponding to gaugeon fields in the effective action.
 Now, such an enlarged action posses both the BRST symmetry as well as 
quantum gauge symmetry. In this enlarged Higgs action the unphysical gaugeon modes are removed by 
more acceptable  Kogo-Ojima type condition. 
 
In this work we have  considered the Yang-Mills theory and Higgs model
to investigate the quantum gauge freedom through Yokoyama gaugeon formalism in the framework of generalized BRST (FFBRST) transformation. We have generalized the BRST symmetry by making the
infinitesimal transformation parameter finite and field-dependent.
Such a generalized BRST transformation is symmetry of the 
action only but not of the generating functional of the Green's functions. 
We  have shown that for a particular finite field-dependent parameter
the Jacobian  of path integral measure under generalized BRST transformation
generates the
gaugeon mode in the effective action in more rigorous way.
We have established the results in both the Yang-Mills and Higgs theories
with explicit calculations.  
Further implications and aspects of present investigations
in certain string theory, M-theory and gravity theory will be more interesting.


\begin{thebibliography}{0}

\bibitem{yo0} K. Yokoyama, Prog. Theor. Phys. 51, 1956 (1974).
\bibitem{yok}  K. Yokoyama,
Prog. Theor. Phys. 59, 1699 (1978).
\bibitem{yoko}  K. Yokoyama,  Prog. Theor. Phys. 60, 1167 (1978);  Phys. Lett. B 79, 79   (1978).
\bibitem{yo1} K. Yokoyama and R. Kubo, Prog. Theor. Phys. 52, 290 (1974).
\bibitem{yo2} K. Yokoyama, M. Takeda and M. Monda, Prog. Theor. Phys. 60, 927 (1978).
\bibitem{yo3} K. Yokoyama, M. Takeda and M. Monda, Prog. Theor. Phys. 64, 1412 (1980).

\bibitem{haya} M. Hayakawa and K Yokoyama, Prog. Theor. Phys. 44,  533 (1970).
\bibitem{ki} K. Izawa, Prog. Theor. Phys. 88, 759 (1992).
\bibitem{mk} M. Koseki, M. Sato and R. Endo, Prog. Theor. Phys. 90, 1111 (1993).
\bibitem{kugo} T. Kugo and I. Ojima, Prog. Theor. Phys. Supplement No. 66,  1 (1979).
\bibitem{kugo1} T. Kugo, I. Ojima, Nucl.Phys. B 144, 234 (1978).

 \bibitem{mk1} M. Koseki, M. Sato and R. Endo,
  Bull. of Yamagata Univ., Nat. Sci. 14, 15 (1996).
\bibitem{naka} Y. Nakawaki,  Prog. Theor. Phys.  98,  5 (1997).
\bibitem{rko} R. Endo  and M. Koseki, Prog. Theor. Phys.  103,    3, (2000).
\bibitem{miu} H. Miura  and R  Endo, Prog. Theor. Phys. 117,  4, (2007).
\bibitem{mir1} M. Faizal, Commun. Theor. Phys. 57, 637 (2012).
\bibitem{mir2} M. Faizal, Mod. Phys. Lett. A 27, 1250147  (2012).


\bibitem{sud0} S. Upadhyay,  EPL 105, 21001 (2014).

\bibitem{sud} S. Upadhyay, Eur. Phys. J. C 74, 2737  (2014);
Annals of Physics 344, 290 (2014).
 
\bibitem{sdj} S. D. Joglekar and B. P. Mandal, Phys. Rev. D 51, 1919 (1995).
\bibitem{sb} S. Upadhyay,   Phys. Lett. B 727, 293 (2013).
\bibitem{sb2} R. Banerjee and S. Upadhyay,  arXiv:1310.1168 [hep-th].
\bibitem{sdj1}  S. D. Joglekar and B. P. Mandal, Int. J. Mod. Phys. A 17, 1279 (2002).
\bibitem{rb} R. Banerjee and B. P. Mandal, Phys. Lett. B 27, 488 (2000).
 \bibitem{susk}   S. Upadhyay,   S. K. Rai and B. P. Mandal,  J. Math. Phys.  {52}, {022301} (2011).
 \bibitem{jog} S. D. Joglekar and A. Misra, Int. J. Mod. Phys. A 15, 1453 (2000).
\bibitem{sb1} S. Upadhyay and B. P. Mandal,  Mod. Phys. Lett.   {A 25}, {3347} (2010);
 EPL 93, 31001 (2011); Eur. Phys. J.  {C 72},  2065 
(2012); Annals of Physics {327}, 2885 (2012);   AIP Conf. Proc. 1444, 213 (2012). 
\bibitem{smm} S. Upadhyay, M. K. Dwivedi and B. P. Mandal, Int. J. Mod. Phys. A 28, 1350033 (2013).
\bibitem{fs} M. Faizal, B. P. Mandal and S. Upadhyay, Phys. Lett. B 721, 159 (2013).
\bibitem{sud1} B. P. Mandal, S. K. Rai and S. Upadhyay, EPL 92, 21001 (2010).
\bibitem{rbs} R. Banerjee, B. Paul and S. Upadhyay, Phys. Rev. D 88, 065019 (2013) .
\bibitem{sudha} S. Upadhyay, EPL  104, 61001  (2013). 
\bibitem{ym} C. N. Yang  and R. L. Mills, Phys. Rev. 96, 191 (1954). 
\bibitem{higg} P. W. Higgs, Phys. Rev. 145, 1156 (1966).
\bibitem{kibble} T. W. B. Kibble, Phys. Rev. 155, 1554 (1967).
\bibitem{fuz} K. Fujikawa and B. W. Lee and A. I. Sanda, Phys. Rev. D 6, 2923 (1972).

\end{thebibliography}
\end{document}